\documentclass[pdflatex,sn-mathphys-num]{sn-jnl}
\usepackage{graphicx}%
\usepackage{multirow}%
\usepackage{amsmath,amssymb,amsfonts}%
\usepackage{amsthm}%
\usepackage{mathrsfs}%
\usepackage[title]{appendix}%
\usepackage{xcolor}%
\usepackage{textcomp}%
\usepackage{manyfoot}%
\usepackage{booktabs}%
\usepackage{algorithm}%
\usepackage{algorithmicx}%
\usepackage{algpseudocode}%
\usepackage{listings}%
\usepackage{subfig}
\algrenewcommand\algorithmicrequire{\textbf{Input:}}
\algrenewcommand\algorithmicensure{\textbf{Output:}}

\begin{document}

\title[Article Title]{Training-Free Query Optimization via LLM-Based Plan Similarity}

\author*[1]{\fnm{Nikita} \sur{Vasilenko}}\email{vasilenko.nikita.research@gmail.com}

\author[1]{\fnm{Alexander} \sur{Demin}}\email{alexandredemin@yandex.ru}

\author[2]{\fnm{Vladimir} \sur{Boorlakov}}\email{vladimir.boorlakov@gmail.com}

\affil[1]{\orgname{Ershov Institute of Informatics Systems}, \orgaddress{\city{Novosibirsk}, \country{Russia}}}

\affil[2]{\orgname{Novosibirsk State Technical University}, \orgaddress{\city{Novosibirsk}, \country{Russia}}}

\abstract{Large language model (LLM) embeddings offer a promising new avenue for database query optimization. In this paper, we explore how pre-trained execution plan embeddings can guide SQL query execution without the need for additional model training. We introduce LLM-PM (LLM-based Plan Mapping), a framework that embeds the default execution plan of a query, finds its k nearest neighbors among previously executed plans, and recommends database hintsets based on neighborhood voting. A lightweight consistency check validates the selected hint, while a fallback mechanism searches the full hint space when needed. Evaluated on the JOB-CEB benchmark using OpenGauss, LLM-PM achieves an average speed-up of 21\% query latency reduction. This work highlights the potential of LLM-powered embeddings to deliver practical improvements in query performance and opens new directions for training-free, embedding-based optimizer guidance systems.}

\keywords{Query Optimization, LLM for Databases, Database Hints}

\maketitle

\section{Introduction}\label{sec:intro}
Query optimization is a core problem in relational database systems: the optimizer must construct an efficient execution plan for each SQL query from a vast search space of alternatives (potential join orders, join algorithms, table access methods, etc.). Decades of research have produced sophisticated cost-based optimizers, yet they remain imperfect and can still choose suboptimal plans that dramatically slow down query execution. A primary cause are inaccurate cardinality estimation and cost model which often relies on simplistic assumptions and thus can be wildly wrong. These mistakes lead to poor plan choices (e.g., using a nested-loop join where a hash join would be far faster).

Modern database systems offer optimizer hints as a mechanism to influence or override plan choices. Hints are directives embedded in the SQL that guide the optimizer to consider specific strategies, such as using a particular join order, favoring a hash join over other join types, or using an index on a given table. In principle, a skilled DBA can use hints to steer the optimizer away from known bad decisions and toward a better plan. For example, a hint could force the use of a previously ignored index, or constrain the join order to avoid inefficient joins. The challenge, however, is that selecting the right hints is extremely complex – it requires a deep understanding of the query, the data, and the optimizer’s internal heuristics. An incorrect hint may even backfire and degrade performance rather than improve it. 

These difficulties have sparked interest in automating query optimization with machine learning. A recent trend is to treat the optimizer as a black box and use learning methods to assist it by suggesting hints or plan tweaks that lead to faster executions. This avoids wholesale replacement of the optimizer, focusing instead on guiding it toward better decisions. Recent studies have shown that Large Language Models (LLMs) can significantly enhance database query optimization. These models are increasingly employed in various ways, such as generating execution plans, embedding queries to understand their semantics, and providing optimization hints. The promising results from these approaches indicate that LLMs are becoming a valuable tool in the field. Given this progress, we believe that further exploration into LLM-driven query optimization is a worthwhile endeavor, with the potential to advance the efficiency and adaptability of database systems. However, despite these encouraging advances, LLM-based techniques remain far from being practical for real-world deployment due to the high cost and latency of inference, which limits their applicability in time-sensitive or resource-constrained environments.

In this paper, we introduce a straightforward method that leverages Large Language Model (LLM) embeddings of execution plans—not the queries themselves—to identify optimal sets of optimizer hints. Our approach combines the representational power of LLM embeddings with a simple, fast, analogy-based prediction system that requires no training.We aim to bridge the gap between the expressive power of LLMs and the need for rapid, efficient query optimization systems. Our method effectively learns from past experiences without requiring an extensive training phase—the heavy lifting is done by the pre-trained LLM, which provides a semantic representation of execution plans.

To validate this approach, we built a prototype on the openGauss database system and evaluated it on one standard benchmark: JOB-CEB (3133 query instances derived from 16 query templates on the IMDB dataset). In our experiments, the LLM-based method successfully guided the optimizer toward faster plans for a large portion of the queries. The results demonstrate that using LLM embeddings for plan hinting can consistently outperform the default optimizer, even without relying on complex training or modifications to the underlying system.

In summary, this paper makes the following contributions:
(1) We introduce a method for query optimization that uses LLM-generated execution plan embeddings to enhance optimizer hint selection.
(2) We propose a lightweight, fast optimization approach that leverages pre-trained LLMs to guide the database optimizer toward better execution plans without requiring extensive training or system modifications.
(3) We demonstrate the effectiveness of LLM-PM through experiments on standard benchmark (JOB-CEB), showing that LLM-based plan hinting can outperform the default optimizer.
(4) We highlight the potential of LLM embeddings to improve query performance, offering a promising direction for efficient, LLM-driven query optimization systems.

\section{Related Work}\label{sec:related}

Recent research into improving the quality of query optimization can be broadly divided into two main directions. The first is aimed at completely replacing the traditional cost-based optimizer with learned models that generate query execution plans from scratch. The second, more incremental direction seeks to augment or guide the existing optimizer using machine learning techniques by refining cost estimates, improving cardinality predictions, or introducing optimizer hints.

\subsection*{Optimizers that \emph{Replace} the Traditional Optimizer}

A first research line seeks to \textbf{replace the entire cost–based optimizer} with a learned model that decides the execution plan end‑to‑end. 

\textbf{Neo} \cite{marcus2019neo} pioneered this direction: a tree‑convolution network predicts the latency of a candidate plan and a search component selects the plan with the lowest predicted latency. This approach demonstrated the feasibility of end-to-end learned planning but required a long training time and extensive labeled data.
Deep‑learning cardinality models can also partially replace core optimizer logic: the MSCN model in \textbf{Learned cardinalities: Estimating
correlated joins with deep learning} \cite{kipf2019learnedcard} and its follow‑up \textbf{Cardinality Estimation with Local Deep Learning Models} \cite{ortiz2019localce} learn complex data correlations that traditional histograms miss, yielding better plans when integrated.  

These works demonstrate that partial or even full replacement of an optimizer is technically feasible, yet they come with significant trade‑offs: (i) they require very large training sets; (ii) they discard decades of handcrafted heuristics embodied in mature optimizers; and (iii) they raise open questions about how well the learned models generalize to unseen data distributions, schemas, and query workloads.

\subsection*{Optimizers that \emph{Guide} the Traditional Optimizer}

A second, more pragmatic line keeps the native optimizer but \textbf{steers} it with learned signals such as cost and cardinality corrections or optimizer hints.

\textbf{AQO} \cite{ivanov2017aqo} is a PostgreSQL extension that
\emph{rewrites only the cardinality-estimation component} of the native
optimizer.  After each query execution, AQO stores the observed
cardinalities for every plan node in a per-query “memory’’ and trains a
$k$-nearest-neighbour regressor in the feature space.  When the same query template re-appears, the learned
estimates are injected in place of the built-in formulas, letting the
standard search procedure pick a (usually much cheaper) plan.  Three
modes (\texttt{learn}, \texttt{use}, \texttt{intelligent}) allow DBAs to
trade warm-up time for safety. Downsides are the
need for several training executions, extra shared-memory to hold the
statistics, and instability in case of data shift.

\textbf{ACM} \cite{vasilenko2024acm} tackles a complementary problem:
\emph{out-of-date cost parameters}.  Implemented inside openGauss, ACM
monitors each query’s buffer-hit ratio and per-operator CPU times, then
uses lightweight linear regression plus exponential smoothing to
continuously retune five key constants of the cost model
(\texttt{seq\_page\_cost}, \texttt{random\_page\_cost}, 
\texttt{cpu\_tuple\_cost}, \texttt{cpu\_operator\_cost} and
\texttt{cpu\_index\_tuple\_cost}).  This online calibration increases
the cost–time correlation for plan nodes and cuts the
end-to-end runtime all without any calibration workload or
changes to the search space.  Because ACM leaves cardinality estimation
untouched, it cannot help when the dominant error source is row-count
misestimation; nonetheless it offers a low-overhead, DBMS-internal way
to keep the cost model in sync with changing hardware and cache
conditions.

\textbf{BAO} \cite{marcus2021bao} frames hint selection as a contextual bandit. The approach uses reinforcement learning to iteratively steer PostgreSQL’s optimizer via a learned hints. Bao continuously learns from running queries, treating the optimizer’s plan choices as a bandit problem to gradually bias it toward better decisions. It achieved substantial speedups over PostgreSQL on benchmark workloads. While powerful such approach tailored to a specific DBMS, instance and data distribution.

To reduce integration effort and generalize across systems, hybrid learned optimizers have been proposed. Anneser et al. (VLDB 2023) present \textbf{AutoSteer} \cite{anneser2023autosteer}, which extends Bao’s framework with automated hint-set discovery and a plug-and-play design that works on multiple SQL engines. AutoSteer treats the optimizer’s tunable “knobs” (e.g., enabling/disabling certain plan rules via hints) as actions and learns which hint combinations to apply for each query to improve performance, without requiring expert-defined hint sets.

\textbf{FASTgres} \cite{woltmann2023fastgres} formulates hint recommendation as a classification task: given a SQL query as input, output a combination of hints (e.g., enable/disable specific join algorithms) that will likely reduce its execution time. Under the hood, FASTgres partitions the query workload into contexts (groups of queries with similar structure) and trains a separate model per context to map query features to hint decisions. The predicted hint set is then applied, and the query runs on the unmodified optimizer. This design treats the optimizer as a black box, and FASTgres effectively learns a function $f:\text{SQL query} \to \text{hints}$ that optimizes performance. A key challenge for such approaches is reliability: ensuring the ML never recommends a harmful hint. FASTgres addresses this by retraining on any query where a predicted hint caused a slowdown (a form of feedback loop).

\textbf{COOOL} \cite{coool2023} casts hint choice as a learning‑to‑rank problem, ordering candidate hinted plans rather than predicting absolute cost.  
\textbf{Lero} \cite{zhu2023lero} also adopts a pair‑wise ranking view: a binary classifier learns which of two plans will be faster, avoiding the need for a calibrated cost model, leveraging the native optimizer to generate
candidate plans and using actual execution feedback to label them. By focusing on pairwise choices, Lero avoids the difficulty of exact cost modeling and instead learns to consistently pick better plans.
Finally, \textbf{HERO} \cite{hero2024} addresses the key challenges in learned hint-based query optimization: reliability, efficiency, and fast inference. Instead of a single neural network, HERO employs an ensemble of context-aware models—each specialized for queries with the same default plan—and organizes them in a graph-based storage of past plans and their performance under various hints. This design ensures interpretable and reliable predictions by avoiding suggestions in unfamiliar contexts. To overcome the exponential search space, HERO uses a parameterized local search algorithm that balances between training cost and optimization quality, enabling rapid convergence to effective hint sets without requiring exhaustive enumeration.

These approaches preserve the DBMS’s tried‑and‑tested search space and heuristics while adding a learned “advisor’’ layer, typically with far lower engineering overhead than wholesale replacement.

\subsection*{Large Language Models (LLMs) for Query‑Plan Optimization}
Recent research has begun to explore whether the rich prior knowledge encoded in foundation models—such as large language models (LLMs)—can be leveraged to improve query optimization.

Tan et al. (VLDB 2025) propose \textbf{LLM‑QO} \cite{tan2025llmqopt}, an optimizer that uses a generative LLM to produce query plans in a textual form. Authors fine-tune a general-purpose LLM with a huge training set of queries and plans (using a technique called QInstruct to serialize database metadata, SQL, and plans into text). LLM-QO treats query optimization as a text generation problem: given the query (and some DB metadata) as input, the LLM directly writes out an execution plan step by step. This is an extreme end-to-end use of an LLM, essentially replacing the optimizer’s search process with the LLM’s learned knowledge.
While this is an exciting and novel direction that showcases the potential of LLMs for direct plan generation, we believe the approach is still far from practical deployment. More extensive studies are needed to convincingly demonstrate that LLMs can consistently produce high-quality execution plans across diverse workloads and schemas. Additionally, the computational cost and latency of generating plans using large language models currently remain significantly higher than those of traditional optimizers, limiting their real-world applicability.

An alternative direction, closer to our work, is using LLMs for plan tuning rather than full plan generation. \textbf{LLMSteer} \cite{akioyamen2024unreasonableeffectivenessllmsquery} is an approach that keeps the existing optimizer but uses an LLM to assist in picking hints. They embed each query’s text using a pre-trained LLM (specifically, they obtain a fixed-size vector embedding of the SQL query), and then train a simple classifier on a small labeled dataset to decide how to steer the optimizer. In their prototype, the task was to choose between two possible plan options for each query (e.g., with or without a particular index or join hint), and the classifier learned to predict which option would be faster. Despite the simplicity of this approach (no complex feature engineering, just raw LLM embeddings as input), it managed to outperform PostgreSQL’s built-in optimizer on the evaluated queries.

In the work \textbf{LLM-R2} \cite{li2024llmr2}, query rewriting is used as an optimization method. This approach solves a similar query optimization task using a language model, but the method differs. The language model is prompted with previously optimized queries and their rewrite rules and tasked with providing recommended rewrite rules for new queries. However, LLM-R2 can only compose rules that Calcite already knows; it cannot invent semantics-preserving rewrites outside that catalogue. 

\paragraph{Position of Our Work.}
Our goal is not to chase state-of-the-art speed-ups at any cost. Instead, we aim to demonstrate both the practical value of execution plan embeddings and the feasibility of building a simple, robust system around them—a system that can realistically be used in practice. LLM-PM is deliberately lightweight and training-free: we use a pre-trained LLM to embed query plans and apply a simple nearest-neighbour search to transfer hint sets. Despite its simplicity, this approach proves to be surprisingly effective—plan embeddings alone can match or even surpass the quality of handcrafted plan encodings, while requiring far less engineering effort.

Crucially, our method avoids the complexity and computational overhead of generative LLM-based approaches. Using only plan embeddings extracted from a frozen LLM is both significantly cheaper and faster than generating plans or rewrite rules with an LLM, making it a far more practical solution for real-world deployment. The nearest-neighbour search component is not only efficient but also robust and transparent—a training-free, plug-and-play mechanism that can easily adapt to new workloads, schemas, and data distributions.

In short, we view our contribution as evidence that "off-the-shelf" plan embeddings, combined with simple retrieval-based methods, offer a practical and low-overhead alternative to both traditional optimizers and complex learned pipelines—not as a bid for theoretical peak performance, but as a step toward real-world usability.

\section{LLM-PM}\label{sec:problem}
This chapter presents LLM-PM, a two-part framework for automatic hint selection based on LLM plan embeddings. It combines our proposed plan-mapping algorithm, which transfers hint sets from similar plans in embedding space, with an adaptive search procedure that exhaustively explores the hint space to identify optimal configurations. While the search algorithm itself is not novel, we include it as a fallback mechanism to ensure completeness and make the overall system self-contained.

\subsection{Problem formulation}
We formalize the task of automatic hint selection for query optimization as follows. Let $Q$ be a SQL query submitted to the database. The database’s query optimizer can produce an execution plan P for $Q$, which results in some cost $C(Q, P)$. In the default case (with no hints), the optimizer chooses a plan $P_{\text{default}}$ based on its built-in heuristics and cost estimates, yielding execution time $T(Q,P_{\text{default}})$. Our goal is to find a set of hints (hintset) $H$ for query $Q$ such that when the optimizer is guided by $H$ (and thus may choose a different plan $P_H$), the execution time is minimized. Formally, we seek:

\begin{equation}
    H^*(Q) = \arg\min_{H \in \mathcal{H}(Q)} \; T\left(Q,\, P_H\right)
\end{equation}

where $\mathcal{H}(Q)$ denotes the space of all valid hint configurations for query $Q$. In other words, $H^*(Q)$ is the optimal hint set that yields the lowest latency for $Q$. This is akin to the definition of steering an optimizer: selecting a hint or set of hints that leads to the fastest plan for the query. In practice, $\mathcal{H}(Q)$ can be very large.

\subsection{Hints}\label{ssec:hint-search}

Exhaustively testing every hint combination quickly becomes infeasible, so we accelerate the search with two techniques.

\paragraph{Timeout threshold (fastest‑so‑far).}
After each plan execution we update a variable
\(
  T_{\min} \leftarrow \min\bigl(T_{\min},\,T(Q,P)\bigr)
\)
(initially $T_{\min}=T(Q,P_{\text{default}})$).  
If any subsequent hinted plan exceeds the \emph{current} $T_{\min}$, we abort
its execution and mark the associated hint set as \emph{sub‑optimal}.  This
prunes expensive plans early.

\paragraph{Plan caching.}
Each distinct physical plan is cached together with its measured latency.
When the optimizer produces a plan already in the cache, we reuse the stored
latency instead of executing the plan again.  This is most effective when many
hints have little or no impact on the chosen plan.

\medskip
\noindent
The seven binary hints under consideration are:

\begin{itemize}
  \item \texttt{enable\_nestloop}
  \item \texttt{enable\_hashjoin}
  \item \texttt{enable\_mergejoin}
  \item \texttt{enable\_seqscan}
  \item \texttt{enable\_indexscan}
  \item \texttt{enable\_indexonlyscan}
  \item \texttt{enable\_bitmapscan}
\end{itemize}
\paragraph{Tie-breaking Among Equivalent Hint Sets.}
Occasionally the same physical plan \(P\) can be generated by more than one hint configuration. For instance,
\[
  H_{1} = (0,1,1,1,0,1,1), \quad
  H_{2} = (0,1,1,1,0,1,0),
\]
produce an identical plan even though they differ only in the final flag (\texttt{enable\_bitmapscan}). Here, the \(i\)-th bit denotes the \(i\)-th Boolean knob listed earlier; a value of \(1\) disables the operator, whereas \(0\) keeps its default, enabled state. To select a unique “canonical’’ hint set we choose the configuration that disables the \emph{fewest} operators: 
\[
    \widehat{H}(P) \;=\; \arg\min_{H : P_H = P} \sum_{i=1}^{7} h_i.
\] %
In the example above, \(H_{2}\) is preferred because it disables one knob fewer than \(H_{1}\). This rule is motivated by two considerations.\emph{(i)} If a plan can be produced whether a knob is on or off, that knob appears to have no influence on the plan’s efficiency, so leaving it enabled avoids unnecessary restrictions. \emph{(ii)} Turning off additional operators compresses the optimizer’s search space and may cause poor performance on queries or data distributions that differ from the training workload. Selecting the most permissive hint set therefore preserves as much of the optimizer’s built‑in expertise as possible while still steering it toward the desired plan.

\paragraph{Adaptive hint search.} %
Algorithm~\ref{alg:hint-search} scans the entire space of $128$ binary hint sets in a single pass while continually tightening an execution‐time threshold.  The query is first executed with all operators enabled to establish an initial timeout $T_{\min}$ and seed the plan–latency cache $\mathcal{C}$.  The remaining configurations are then dispatched across $w=8$ worker threads.  For each hint set the optimizer produces a physical plan; if the same plan has already been observed, its latency is retrieved from $\mathcal{C}$ without re‐execution.  Otherwise the query is executed under the current timeout $T_{\min}$.  Plans that exceed the timeout are aborted early and the corresponding hint sets are labelled \emph{sub-optimal}. Whenever a faster execution finishes, $T_{\min}$ is atomically updated, immediately tightening the timeout for all workers still in flight. Because slow plans are terminated quickly and repeated plans are short-circuited via the cache, the procedure spends most of its effort on a small number of promising configurations while still guaranteeing full coverage of the hint space.

\begin{algorithm}[H]
\caption{Adaptive Hint Search with Timeout Pruning and Plan Cache}
\label{alg:hint-search}
\begin{algorithmic}[1]                 

\Require Query $Q$;
        full catalogue of binary hint sets
        $\mathcal{H}=\{H_1,\dots,H_{128}\}$;
        worker count $w \gets 8$
\Ensure  runtime threshold $T_{\min}$;
        plan–latency cache $\mathcal{C}$

\State $T_{\min} \gets \textsc{execute}(Q,\text{default})$
\State $\mathcal{C} \gets \bigl\{\bigl(\textsc{plan}(Q,\text{default}),\,T_{\min}\bigr)\bigr\}$

\ForAll{$H \in \mathcal{H}\setminus\{\text{default}\}$ \textbf{in parallel on} $w$ threads}
    \State $P \gets \textsc{plan}(Q,H)$
    \If{$P \in \mathcal{C}$}
        \State $t \gets \mathcal{C}[P]$
    \Else
        \State $t \gets \textsc{execute}(Q,H,\text{timeout}=T_{\min})$
        \If{$t = \textsc{Timeout}$}
            \State mark $H$ as \emph{sub-optimal}
            \State \textbf{continue}
        \EndIf
        \State $\mathcal{C}[P] \gets t$
    \EndIf
    \State $T_{\min} \gets \min\!\bigl(T_{\min},\,t\bigr)$ \Comment{atomic update}
\EndFor

\State \Return $(T_{\min},\,\mathcal{C})$
\end{algorithmic}
\end{algorithm}

\subsection{Plan-Mapping Algorithm}\label{ssec:mapping-algo}
This section will describe the principle of plan mapping and its motivation and  how a new query plan receives hints from the existing embedding space.
Based on the assumption that embeddings from an LLM can provide a good representation of a query plan, we can attempt to build a system that assigns the same hint set to similar plans. To achieve this, it is necessary to define a distance metric between plans, where plans within a certain distance from each other would be assigned the same hint set.

When planning a query with a new hint, a new, potentially faster plan emerges. It would be useful to also consider this new plan in the system. Following this idea, a second embedding for the potentially faster plan can be created and compared with the optimized plan resulting from applying the hint in the embedding store. In this way, the assignment of hints will undergo a two-step verification process: first, comparing the default plans and searching for a hint candidate, and second, evaluating the modified plan. 

The goal is to decide, for a \emph{new plan} $P_0$ (obtained for the
incoming query $Q$ with all hints enabled), whether there exists a nearby plan
in the embedding space whose associated hint set is likely to improve~$Q$.
We operate in the space of plan embeddings produced by a pre-trained LLM; the
word “plan’’ below always refers to its embedding.

From the offline search described in \S\ref{ssec:hint-search} we store triples
\(
  (\mathbf{d}_i,\,H_i^\star,\,\mathbf{o}_i)
\),
where $\mathbf{d}_i$ is the embedding of the $i$-th \emph{default} plan,
$H_i^\star$ is its optimal hint set, and $\mathbf{o}_i$ is the embedding of the
corresponding \emph{optimised} plan.  If no hint set speeds up the query, then
$H_i^\star$ is the all-enabled vector and $\mathbf{o}_i=\mathbf{d}_i$.

\paragraph{Key intuition.}
Plans that are close in embedding space tend to share physical structure;
therefore the same hint set often remains beneficial within a local
neighbourhood.  We leverage this locality in two stages:

\begin{enumerate}
    \item \textbf{Neighbourhood voting.} Among the 
    $N$ nearest default plans, select the most popular non-default hint set. This rests on the assumption that structurally similar plans tend to share the same inefficiencies—so applying the hint that worked for those neighbors is likely to optimize the new plan in a similar way.
    \item \textbf{Consistency check.} Re-plan the query with the chosen candidate hint set and for both the default plan and the candidate-hint plan we gather their $K$ nearest neighbors (all of which have known execution times), compute the average runtime in each neighborhood, and compare them. If the candidate-plan neighborhood’s average runtime is smaller than the default-plan neighborhood’s, we adopt the new hint set; otherwise, we keep the original.
\end{enumerate}

This two-stage test balances coverage and safety: the $N$ nearest neighbours let us compile a list of locally popular hint sets and select the most frequent one as the candidate, while the consistency check with $K$ prevents us from blindly applying a hint that would push the optimizer to sub-optimal space. 

\begin{figure}[ht]
  \centering
  \includegraphics[width=0.8\textwidth]{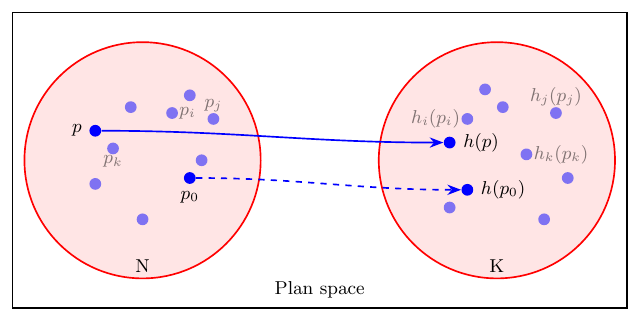}
  \caption{Two-stage hint selection: vote among the $N$ nearest default-plan neighbours to choose a candidate hint, then verify its benefit via the $K$-nearest optimized-plan neighbourhood.}
  \label{fig:plan‐mapping}
\end{figure}

\paragraph{Detailed description.}
Let $\phi(\cdot)$ be the LLM encoder mapping a query plan to a $d$-dimensional vector.  In Figure~\ref{fig:plan‐mapping}, the left circle  shows the embedding of the default plan $P_0$ as point $p_0$, together with its $N$ nearest neighbours (semi-transparent blue).  We vote among those neighbours’ hint sets to pick the candidate hint $H_{\mathrm{cand}}$.  On the right, we show the embedding of the re-planned query $P_{\mathrm{cand}}$ as $h(p_{0})$ and its $K$ nearest neighbours.  By comparing the average runtimes in these two $K$-neighbourhoods—one around $p$ (using default runtimes) and one around $p_{\mathrm{new}}$ (using optimized runtimes)—we decide whether $H_{\mathrm{cand}}$ genuinely improves performance.  This two-stage check maximizes hint coverage (via the first $N$-neighborhood vote) while guarding against harmful suggestions (via the second $K$-neighborhood consistency test).

\begin{enumerate}
  \item \textbf{Embed the default plan.}  
    Compute
    \[
      \mathbf{p}_0 = \phi(P_0),
    \]
    where $P_0$ is the optimizer’s default plan for query $Q$.

  \item \textbf{Neighbourhood voting.}  
    Find the $N$ nearest neighbours of $\mathbf{p}_0$:
    \[
      \mathcal{N}
      = \bigl\{(\mathbf{d}_i, H_i^\star, t_i)
               \mid i\in\text{top-}N\text{ closest to }\mathbf{p}_0
         \bigr\},
    \]
    where each neighbour carries its hint set $H_i^\star$ and known runtime $t_i$.  
    Build a frequency table over non-default $\{H_i^\star\}$ and choose the most frequent as $H_{\mathrm{cand}}$, breaking ties by smallest distance $\|\mathbf{p}_0-\mathbf{d}_i\|$.

  \item \textbf{Embed the candidate plan.}  
    Apply $H_{\mathrm{cand}}$ to $Q$, let the resulting plan be $P_{\mathrm{cand}}$, and compute
    \[
      \mathbf{p}_{\mathrm{cand}} = \phi(P_{\mathrm{cand}}).
    \]

  \item \textbf{Consistency check via two $K$–neighborhoods.}  
    \begin{itemize}
      \item Let 
        \[
          \mathcal{K}_0
          = \{\mathbf{d}_{0,1}, \dots, \mathbf{d}_{0,K}\}
          \quad\text{and}\quad
          \mathcal{K}_{\mathrm{cand}}
          = \{\mathbf{d}_{\mathrm{cand},1}, \dots, \mathbf{d}_{\mathrm{cand},K}\}
        \]
        be the $K$ nearest neighbours of $\mathbf{p}_0$ and $\mathbf{p}_{\mathrm{cand}}$, respectively.
      \item Retrieve their known runtimes
        $\{t_{0,1},\dots,t_{0,K}\}$ and $\{t_{\mathrm{cand},1},\dots,t_{\mathrm{cand},K}\}$.
      \item Compute the average runtimes
        \[
          \bar t_0
          = \frac1K\sum_{j=1}^K t_{0,j},
          \qquad
          \bar t_{\mathrm{cand}}
          = \frac1K\sum_{j=1}^K t_{\mathrm{cand},j}.
        \]
      \item \emph{Accept} $H_{\mathrm{cand}}$ if
        $\bar t_{\mathrm{cand}} < \bar t_0$;
        otherwise, \emph{reject} it (i.e.\ keep the default plan).
    \end{itemize}
\end{enumerate}

\paragraph{Hyper-parameters.}
The algorithm has three knobs:

\begin{itemize}
  \item $N$ – neighbourhood size for the voting stage (\(N\!=\!16\) in our
        experiments);
  \item $K$ – consistency check via two K–neighborhoods; small $K$ favours closest plans, large $K$
        increases coverage and stability (\(K\!=\!16\) also in our
        experiments);
  \item distance metric (euclidean by default).
\end{itemize}

\begin{algorithm}[H]
\caption{Plan-Mapping Algorithm with Two-Neighborhood Consistency}
\label{alg:plan-mapping2}
\footnotesize                             
\begin{algorithmic}[1]                    

\Require new query $Q$;
        reference set
        $\displaystyle
          \mathcal{T}=\{(d_i, H_i^\ast, o_i,
          \,t_i^{\mathrm{def}},\,t_i^{\mathrm{opt}})\}_{i=1}^{M}
        $;
        parameters $N$ (voting-neighbours), $K$ (consistency-neighbours)
\Ensure  hint set $H_{\mathrm{out}}$ \textbf{or} $\varnothing$

\State $P_0 \gets \textsc{plan}\!\bigl(Q,\text{default hints}\bigr)$
\State $\mathbf p_0 \gets \phi(P_0)$
\State $\mathcal{N} \gets$ the $N$ nearest $\{d_i\}$ to $\mathbf p_0$
\vspace{2pt}

\State $H_{\mathrm{cand}} \gets
       \operatorname{mode}\!\bigl\{\,H_i^\ast \mid
       H_i^\ast \neq \mathbf 0,\; (d_i,\dots)\in\mathcal{N}\bigr\}$
\If{$H_{\mathrm{cand}}$ is undefined}
    \State \Return $\varnothing$
\EndIf
\vspace{2pt}

\State $P_{\mathrm{cand}} \gets \textsc{plan}(Q,H_{\mathrm{cand}})$
\State $\mathbf p_{\mathrm{cand}} \gets \phi(P_{\mathrm{cand}})$

\State $\mathcal{K}_0 \gets$ the $K$ nearest $\{d_i\}$ to $\mathbf p_0$
\State $\displaystyle
       \bar t_0 \gets \frac1K
       \sum_{(d_i,\dots,t_i^{\mathrm{def}},\dots)\in\mathcal{K}_0}
       t_i^{\mathrm{def}}
       $
\State $\mathcal{K}_{\mathrm{cand}} \gets$ 
       the $K$ nearest $\{o_i\}$ to $\mathbf p_{\mathrm{cand}}$
\State $\displaystyle
       \bar t_{\mathrm{cand}} \gets \frac1K
       \sum_{(o_i,\dots,t_i^{\mathrm{opt}},\dots)\in\mathcal{K}_{\mathrm{cand}}}
       t_i^{\mathrm{opt}}
       $
\vspace{2pt}

\If{$\bar t_{\mathrm{cand}} < \bar t_0$}
    \State $H_{\mathrm{out}} \gets H_{\mathrm{cand}}$
\Else
    \State $H_{\mathrm{out}} \gets \varnothing$
\EndIf
\State \Return $H_{\mathrm{out}}$
\end{algorithmic}
\end{algorithm}

\subsection{Overall system design}\label{ssec:overall-system}

\begin{figure}[htbp]
  \centering
  \includegraphics[width=.85\linewidth]{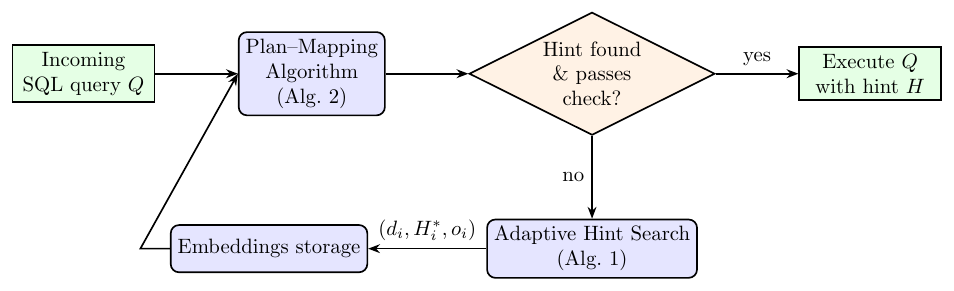}
  \caption{Overall-system control flow of LLM-PM.}
  \label{fig:overall-system}
\end{figure}

To effectively combine the strengths of our embedding-based approach, we integrate all proposed techniques into a unified query optimization pipeline. This hybrid design is intended to (1) enable fast query optimization via Plan-Mapping when similar past plans exist in the experience store, (2) fall back to a thorough hint search for novel queries, and (3) improve overall performance over time through continual accumulation of query execution experience. Figure~\ref{fig:overall-system} depicts the two-stage optimization pipeline.

\begin{enumerate}
  \item \textbf{Plan-Mapping stage}  
        The incoming SQL query~$Q$ is planned with all hints
        enabled, embedded, and compared to the reference workload
        using Algorithm~\ref{alg:plan-mapping2}.  
        If the selected candidate hint~$H_{\mathrm{cand}}$
        passes the two-neighbourhood consistency test, the query proceeds
        directly to execution with $H_{\mathrm{cand}}$ (“fast path’’).

  \item \textbf{Fallback to Adaptive Hint Search}  
        When Plan-Mapping returns \emph{no} hint (or the consistency test
        rejects the candidate), $Q$ is forwarded to
        Algorithm~\ref{alg:hint-search}.  
        This exhaustive yet timeout-pruned search
        inspects the entire space of 128 binary hint sets, finds the
        fastest configuration, and caches optimal observed
        \(\langle\text{default plan},\text{optimal hint}, \text{optimal plan}\rangle\) triplet for future use.

  \item \textbf{Execution and continual learning.}  
        Triplets produced by the fallback search continuously
        enlarge the reference set, which the Plan-Mapping stage
        consults on every new query (feedback arrow from
        \textsf{store} to \textsf{map}).  
        Over time this feedback loop shrinks the fraction of
        queries that require the slower exhaustive search,
        steadily improving end-to-end latency.
\end{enumerate}

This design yields low median latency—because most queries are solved by the
lightweight, embedding-based mapper—while guaranteeing that even “difficult’’
queries eventually benefit from the thorough search routine.

\section{Evaluation}
In this chapter we first outline the experimental design—including datasets, baselines, and implementation details—before detailing the quantitative and qualitative metrics used to assess performance. We then present the results, analyze their statistical significance, and discuss the practical implications that emerge. Finally, we highlight the key insights learned  from the evaluation, setting the stage for the conclusions that follow.
\subsection{Setup and dataset}

\begin{figure}[htbp]
  \centering
  \includegraphics[width=0.5\linewidth]{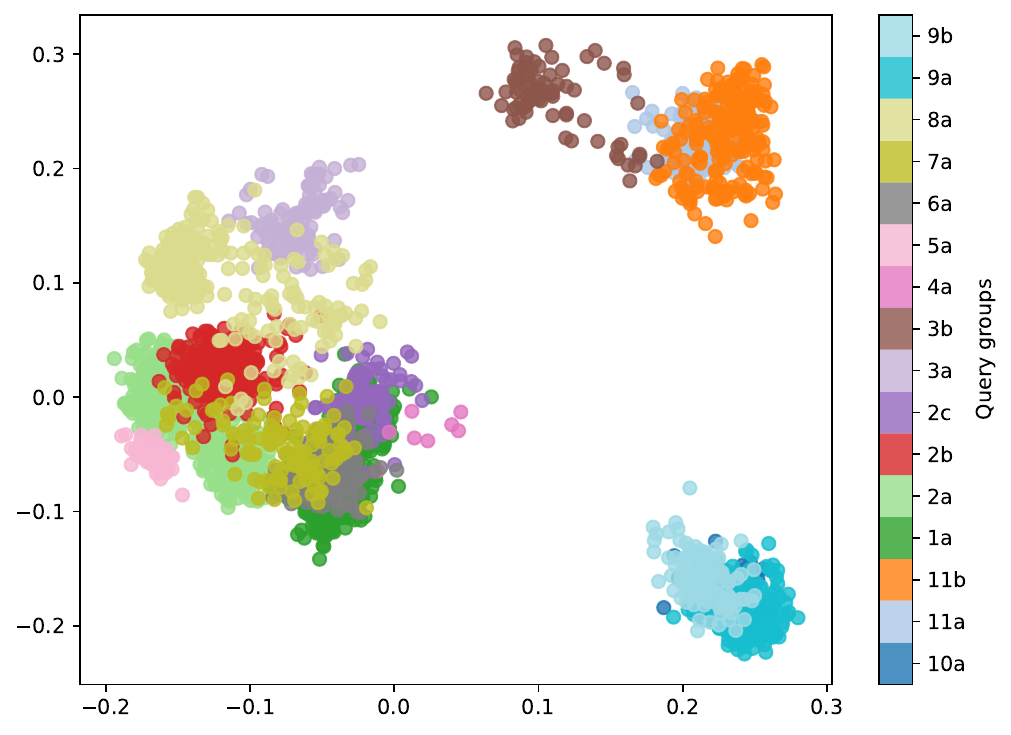} 
  \caption{Default plans after PCA}
  \label{fig:pca_scatter}
\end{figure}

The experiments were conducted on a server with 128 CPU cores based on HiSilicon Kunpeng-920 processors, with a total of 1000 GB of RAM. Storage was provided by a 2 TB SSD NVMe disk, offering high read/write speeds for fast data access. The database used was single-node OpenGauss. The following configuration settings were applied:
\begin{itemize}
    \item \texttt{shared\_buffers = 256 GB}
    \item \texttt{bulk\_write\_ring\_size = 10 GB}
    \item \texttt{work\_mem = 80 GB}
    \item \texttt{cstore\_buffers = 4 GB}
    \item \texttt{query\_dop = 1} (single-threaded query execution for optimal control over performance)
\end{itemize}

The experiments were carried out on the IMDB dataset, which contains 3 133 queries drawn from the JOB-CEB workload. The dataset relies on 16 templates from the original JOB benchmark, and each template contributes a different number of queries. Plan embeddings were generated with OpenAI’s text-embedding-3-large model, so every query is paired with both its default and optimized execution plans, their embeddings, and the associated runtimes.
Figure \ref{fig:pca_scatter} illustrates how the default plans are arranged after reducing the high-dimensional embeddings to two dimensions with Principal Component Analysis (PCA). Each point denotes a single plan projected onto the first two principal components, and its color encodes the corresponding query-group label shown in the legend.

Previous studies on PostgreSQL report that a handful of individual hints
can by themselves accelerate the JOB--CEB workload by more than
60 \%; for example, globally disabling nested-loop joins yields such a
speed-up.  In contrast, turning off the same join operator in
\emph{openGauss} degrades performance by roughly 50 \%, substantially
complicating the task of selecting an optimal hint set.

\begin{figure}[htbp]
  \centering
  \begin{minipage}[t]{0.48\linewidth}
    \centering
    \textbf{(a)}\\[4pt]               
    \small                            
    \begin{tabular}{l r}
      \toprule
      Hintset & Count \\
      \midrule
      0000000 & 1468 \\
      0001100 & 226  \\
      1000100 & 146  \\
      1100100 & 114  \\
      0000100 & 106  \\
      0100100 & 46   \\
      0100000 & 45   \\
      0000001 & 43   \\
      1000000 & 41   \\
      0011100 & 40   \\
      \bottomrule
    \end{tabular}
  \end{minipage}%
  \hfill                              
  \begin{minipage}[t]{0.48\linewidth}
    \centering
    \textbf{(b)}\\[4pt]
    \includegraphics[width=\linewidth]{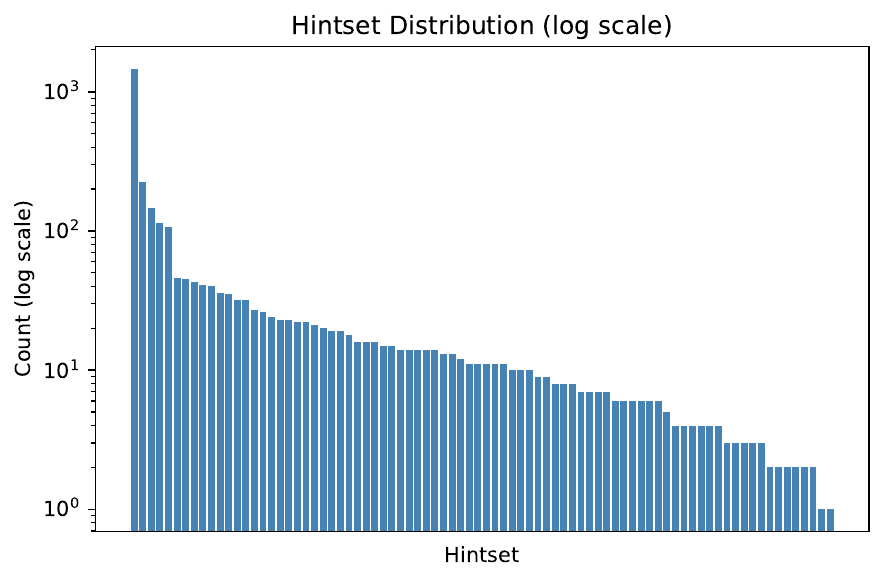}
  \end{minipage}

  \caption{Hint-set statistics: (a) count table and (b) log-scale histogram.}
  \label{fig:hintset_combo}
\end{figure}

Table \ref{fig:hintset_combo}\,(a) lists the ten most frequent
\emph{optimal} hint sets.
For roughly half of the queries the optimal configuration
coincides with the default (all hints enabled);
in total, 86 of the 128 possible bit-vectors were selected as
optimal at least once.
Figure \ref{fig:hintset_combo}\,(b) shows the global distribution of all
observed hint sets.
The leaders are diverse— only disabling \texttt{enable\_nestloop} is
\emph{not} always the winning strategy.

\subsection{Results}\label{ssec:result}

\begin{figure}[htbp]
  \centering
  \begin{minipage}[t]{0.51\linewidth}
    \centering
    \textbf{(a)}\\[4pt] 
    \includegraphics[width=1.1\linewidth]{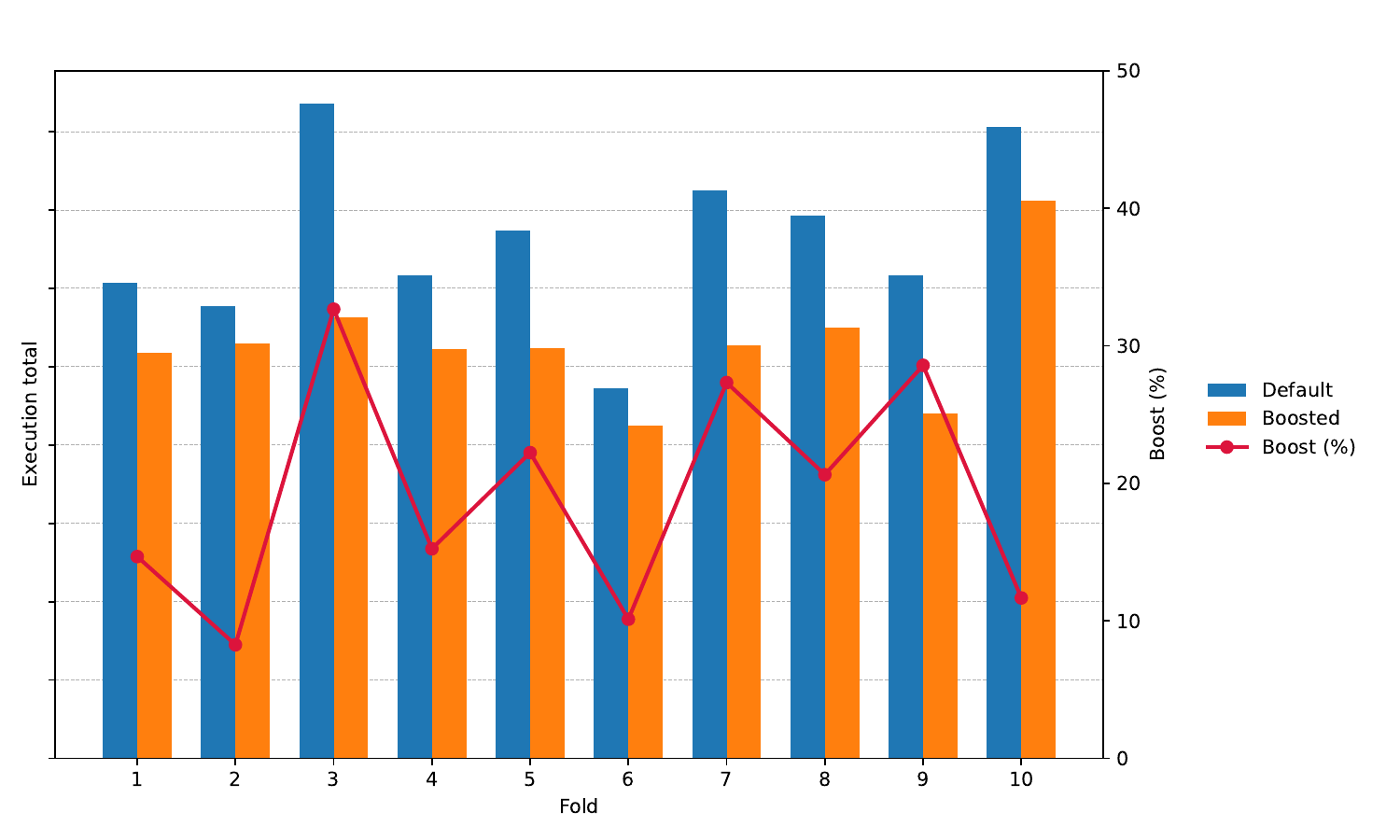}
  \end{minipage}%
  \hfill
  \begin{minipage}[t]{0.48\linewidth}
    \centering
    \textbf{(b)}\\[4pt] 
    \footnotesize
    \begingroup                             
      \setlength{\tabcolsep}{5pt}           
      \renewcommand{\arraystretch}{1.0}     

      \begin{tabular}{l@{\hspace{5pt}}r@{\hspace{5pt}}r@{\hspace{5pt}}r}
        \toprule
        \shortstack[l]{Upper\\percentile\\bound}
          & \shortstack[c]{Mean\\default\\time}
          & \shortstack[c]{Mean\\boosted\\time}
          & Boost \\
        \midrule
        0.1 &   564   &   628   & -11 \\
        0.2 &  1658   &  1651   &   0 \\
        0.3 &  2812   &  2740   &   3 \\
        0.4 &  4203   &  4096   &   3 \\
        0.5 &  6144   &  5954   &   3 \\
        0.6 &  8617   &  8293   &   4 \\
        0.7 & 12802   & 11681   &   9 \\
        0.8 & 19809   & 17059   &  14 \\
        0.9 & 35201   & 27923   &  21 \\
        1.0 & 121154  & 90734   &  25 \\
        \bottomrule
      \end{tabular}
    \endgroup
  \end{minipage}

  \caption{(a) Per‐fold latency improvement under Plan‐Mapping.
           (b) Percentile‐level breakdown of mean runtime and boost (\%).}
  \label{fig:combined}
\end{figure}

Figure~\ref{fig:combined}\,(a) summarizes the outcome of a ten‐fold cross‐validation: in each fold, 10 \% of the JOB‐CEB queries were held out for testing, while the remaining 90 \% seeded the reference store for Plan‐Mapping.  

The aggregate runtime across the ten folds was, on average, reduced by \textbf{19.1 \%} (i.e.\ queries run on average \(1.19\times\) faster than the default optimizer).  This gain is robust: even the slowest fold achieves an \(\,+8\,\%\) speed‐up, and the best fold \(\,+32\,\%\). The total aggregate speed-up achieved—computed by summing the runtimes of all test queries—is 21.1 \%. The maximum achievable speed-up for the entire workload is 62.5 \%. On an \emph{average fold}, 20 \% of queries accelerate, 20 \% decelerate, and the remaining 60 \% are unaffected.  The 90th‐percentile execution time falls by 24.7 \%, while the median time improves slightly by 2.1 \%.

Table ~\ref{fig:combined}\,(b) breaks average query runtime into percentile bands.
The “Upper bound for percentile” column gives the upper edge of each band; the lower edge is the previous percentile. For example, the 0.5 row covers the 0.4–0.5 percentile slice.
“Default” and “Boosted” columns report the mean runtime inside each slice, while “Boost” shows the percentage change within that slice.
Plan-mapping delivers its largest gains in the long-tail slices with the slowest queries, although a few low-latency slices show occasional slow-downs.

\subsection{Ablation study}

A natural question is whether our learned LLM embeddings provide any real benefit beyond acting as an expensive form of string matching. To answer it, we reran the evaluation under two ablated configurations in which the consistency-check stage was disabled, in order to eliminate the influence of the additional verification step and assess whether LLM plan embeddings are truly useful for evaluating plan similarity compared to simple string-based matching metrics:

\begin{enumerate}
  \item \textbf{Levenshtein distance} – candidates are retrieved purely by the Levenshtein edit distance;
  \item \textbf{LLM embeddings (no consistency check)} – identical to our full system except that the consistency check is omitted.
\end{enumerate}

In the Levenshtein variant we no longer embed plans or compare vectors with the Euclidean metric; instead we compute the edit distance
\[
d_{\text{lev}}(p_{1},p_{2})=\min_{\pi}\lvert\pi\rvert ,
\]
where \(\pi\) is a sequence of single-character insertions, deletions, or substitutions that transforms plan string \(p_{1}\) into plan string \(p_{2}\).

Table \ref{tab:ablation-latency} reports the total speed-up and the number of queries that exceeded the 450s time-out.  
Without the consistency check, the Levenshtein variant not only fails to accelerate the workload: it slows it down by \textbf{64.4 \%} and triggers \textbf{87} time-outs.  
By contrast, LLM embeddings still deliver a \textbf{16.1 \%} speed-up with only \textbf{5} time-outs—five times fewer than Levenshtein and close to the default optimizer (3). 
When the consistency check is re-enabled (Section \ref{ssec:result}), the speed-up rises to \textbf{21.1 \%} while time-outs drop to a single query.

These observations confirm that the learned embeddings capture information well beyond raw string similarity and, together with the consistency check, are essential for robust performance gains.

\begin{table}[htbp]
  \centering

  \begin{tabular}{lccc}
    \toprule
                     & Hint selection only
                     & Time-outs \\
    \midrule
    Levenshtein      & $-64.4\%\;$
                     & 87  \\[2pt]
    LLM embeddings   & $16.1\%\;$
                     & 5  \\
    \bottomrule
  \end{tabular}

  \caption{Total latency change (positive = speed-up) and
           number of time-outs for each distance metric}
    \label{tab:ablation-latency}
\end{table}

\subsection{Discussion and future work}

Our ablation study confirmed that learned LLM embeddings are valuable for hint prediction, yet even state-of-the-art OpenAI embeddings are trained on very little material that combines query plans with explicit information on how hints affect them.  A natural next step is therefore to develop or fine-tune domain-specific embeddings whose training corpora contain not only rich SQL syntax but also detailed optimization techniques, hint annotations, and cost feedback.  Such embeddings could underpin more powerful retrieval schemes and, more broadly, improve an LLM’s ability to reason about the inherently complex task of database query optimization.

While the current Plan-Mapping pipeline delivers an average \(\approx 20\%\) acceleration—roughly one-third of the theoretical maximum on this workload—it also exposes clear opportunities for refinement. Half of the remaining gap arises during the candidate‐hint selection stage: in an oracle configuration, where the consistency check is allowed to inspect the true runtimes of both default and candidate plans, we recoup almost \(40\%\) of the attainable speed-up (two-thirds of the maximum).  This suggests that a more sophisticated similarity metric or voting mechanism for the first stage could close much of the residual gap.

Hardware and engine heterogeneity further complicate direct comparisons.  Previous work on PostgreSQL reported over \(60\%\) average speed-up from a single “disable nested-loops” hint, yet on our openGauss platform the same hint regresses performance by more than \(50\%\).  This observation motivates our multi-neighbourhood strategy and highlights that no single hint is universally optimal.

\section{Conclusion}

We have presented LLM-PM, a lightweight, training‐free framework that leverages pre‐trained LLM embeddings of execution plans to steer a traditional cost‐based optimizer via hint recommendations. One of the primary goals of this study was to evaluate the utility of these embeddings, and our experiments provide clear evidence of their effectiveness. Our method first identifies a locally popular hint set among the $N$ nearest default plans, then verifies its benefit through a two‐neighbourhood consistency check that compares average runtimes in both default and optimized plan regions. Importantly, this approach requires no changes to the underlying DBMS or extensive offline training: it simply reuses past plan outcomes encoded in the embedding space.

We evaluated the core component of the LLM-PM system—Plan-Mapping—on openGauss using standard benchmarks (JOB-CEB on the IMDB dataset), demonstrating its consistent improvements in query performance. Looking forward, promising extensions include refining the embedding space with metric learning or hybrid plan‐SQL features and dynamically adapting neighbour sizes based on local density. These enhancements could further narrow the gap toward optimal hint selection while preserving the simplicity and practicality that make Plan-Mapping suitable for real-world deployment.

\bibliography{related_work}

\end{document}